\documentclass[superscriptaddress,secnumarabic,twocolumn,
 amssymb,amsmath,nobibnotes,aps,prd,showkeys,showpacs,nofootinbib]{revtex4}

\RequirePackage{graphicx}
\RequirePackage{mathptmx}   
\RequirePackage{flushend}

\usepackage{mathrsfs}
\usepackage{color}

\DeclareMathOperator{\arccosh}{arcCosh}
\DeclareMathOperator{\arcsinh}{arcsinh}
\DeclareMathOperator{\arctanh}{arctanh}

\usepackage[normalem]{ulem}
\begin{document}

\title{Uniformly accelerated traveler in an FLRW universe}

\author{Morteza Kerachian}
\email{kerachian.morteza@gmail.com}
\affiliation{Institute of Theoretical Physics, Faculty of Mathematics and Physics,
Charles University, CZ-180 00 Prague, Czech Republic}

\begin{abstract}
 This paper introduces an analytical treatment of accelerated and geodesic motion within the framework of the Friedmann -Lema{\^\i}tre-Robertson-Walker (FLRW) spacetime. By employing conformal time transformations we manage to convert second order differential equations of motion in FLRW spacetime to first order equations in the conformally transformed spacetime. This allows us to derive a general analytical solution in closed-form for accelerated motion in spatially curved FLRW spacetime. We provide few examples of this general solution. The last part of our work focuses on the return journey for a traveler exploring a FLRW universe. We derive certain conditions for a de Sitter universe that have to be satisfied in order to achieve a return journey.
\end{abstract}

\pacs{~~}

\maketitle

\section{Introduction}
\label{intro}
The paradigm of a homogeneous expanding isotropic universe  in the framework of General Relativity is realised via the Friedmann-Lema{\^\i}tre-Robertson-Walker (FLRW) model \cite{friedmann1922krummung,lemaitre1931expansion,robertson1935kinematics,walker1936ag}. In this work we are going to investigate the accelerated motion of a test particle in FLRW. Such a test particle corresponds to a rocketeer traveling in an FRLW universe. The derivation and the interpretation of accelerated motion have suffered from ambiguous treatments which will be discussed later on\footnote{For example, it has been debated whether analysing uniformly accelerated motion in an expanding universe could clarify the physics behind the expansion. Namely, the debate has been about if the expansion is a trick of coordinates or a physical phenomenon~\cite{abramowicz2006eppur,abramowicz2008spacetime,abramowicz2008short,lewis2008cosmologicalacc,kwan2010adventures}.}.

The motion of a uniformly accelerated traveller in an expanding universe is described by a set of differential equations which are in general non-trivial coupled. This set of equations does not become less complicated even if a specific cosmological model is applied. Thus, an analytical derivation of the path of a rocketeer is highly challenging. Actually a goal of this work is to present a general formulation which allows an  analytical treatment. In particular, this work has been inspired by \cite{rindler1960hyperbolic} in which W.~Rindler proposed a generalization of the hyperbolic motion in Minkowski spacetime to solve the corresponding set of equations. However, Rindler solved it only for the de Sitter spacetime \cite{rindler1960hyperbolic}. 

Studying the accelerated motion of a rocketeer is useful for the future accelerated space probe. For our universe ( with $\Omega_m\approx 0.27$,  $\Omega_\Lambda\approx 0.73$ and nearly spatially flat) a space traveler could visit a galaxy which is observed today at a redshift of $1.7$ on a one-way journey with proper acceleration equal to the terrestrial gravitational acceleration, in almost 100 years~\cite{heyl2005long}. However, for galaxies at redshift less than $1.7$, e.g. $0.65$, it is not clear whether the traveller would succeed to return back home. Therefore, it might be appropriate to consider a traveller of intermittent accelerations to explore the universe~\cite{kwan2010adventures}. In this study, we are going to address this issue from a different point of view, i.e. our rocketeer travels with uniform deceleration in order to achieve a return trip.  

The formalism presented in this work reduces to the geodesic motion in a spatially curved FLRW spacetime in the limit of zero proper acceleration. Since geodesic in an expanding universe has vast applications in cosmology, astrophysics and quantum gravity, many attempts have been undertaken to solve the geodesic equations of motion (for more details see references $[12-39]$ of \cite{cunningham2017exact}). The first attempt to tackle this issue was initiated by Whiting~\cite{whiting2004expansion}. Whiting derived the equations of motion for a free particle with Newtonian background and its relativistic generalization. In~\cite{davis2003solutions} geodesic in low-velocity regime has been studied. These efforts  by~\cite{whiting2004expansion,davis2003solutions} for solving geodesic motion was not sufficient due to the number of shortcomings in calculation and interpretation. Latter on,  Gr\o n \& Elgar\o y~\cite{gron2007space} derived a general solution for geodesics in the full general relativity framework. Moreover, Ref.~\cite{whiting2004expansion} claimed that particle moving uniformly in an expanding universe will join the Hubble flow. This claim has been refuted in~\cite{barnes2006joining}, in which it has been formally proven that particles following the geodesic motion in an eternally expanding universe do not asymptotically rejoin the Hubble flow. Recently, a method for deriving both timelike and spacelike geodesic distances in spatially flat FLRW spacetime with given initial-value or boundary-value constraints was presented in~\cite{cunningham2017exact}.

In this work, we use conformal time transformations in order to get a general analytical formulation. Thus, it is useful to provide a brief overview of what has been already done in FLRW with conformal transformations. Conformal transformation and its symmetries help us to grasp the notion of the causal structure of spacetime \cite{hawking1973large}. FLRW metric has vanishing Weyl tensor, therefore, all Friedmann cosmological models are conformally flat and their systematic description has been studied in detail in Ref.~\cite{gron2011frw,gron2011frw1,Grn2011}. The nature of FLRW models in conformal coordinates has been studied in~\cite{chodorowski2007direct}. It has been demonstrated in~\cite{lewis2007coordinate} that transformation into conformal coordinates do not eliminate superluminal recession velocities for open or flat matter dominated FLRW cosmologies, and all of them possess superluminal expansion. Ref.~\cite{bivcak2001j} derived the scalar field and the electromagnetic field of a moving charged particle in de Sitter spacetime, when the particle is following geodesic trajectories or it is uniformly accelerated. In order to achieve this, conformal transformation between de Sitter and Minkowski spacetime was applied. 

The layout of this work is as follows; Sec.~\ref{sec2} provides the essential mathematical background in which the conformal time transformation is applied. In Sec.~\ref{sec3}  a novel general formalism is presented. Namely, using the transformed FLRW spacetime enables us to solve the equations of motion of accelerated particle. In this way, second order differential equations reduce to first order differential equations which allow us to solve the trajectories for accelerated particle and free motion. In addition, this formalism specifies the four-velocities of particles. This extends previous results \cite{gron2007space,cunningham2017exact} covering only geodesic motion. We prove that accelerated and geodesic motions in FLRW universe depend on the expansion factor and its integral for any specific FLRW model. In Sec.~\ref{sec4} we give some examples for known FLRW models that have an analytic solution. In cases where there is no analytical solutions, we use numerical integration to solve them. Furthermore, in Sec.~\ref{sec:antidess} we provide solutions for both the uniformly accelerated and the geodesic motion in the global anti-de Sitter spacetime by implementing similar prescription as we did in Sec.~\ref{sec3}. This accelerated trajectory is indeed the generalized form of the known uniformly accelerated observer in the anti-de Sitter spacetime~\cite{griffiths2009exact,podolsky2002accelerating,krtouvs2005accelerated}. In Sec.~\ref{sec:return} we discuss the return journey. We show that, in order to achieve a return journey having uniformly deceleration is not sufficient condition for every spacetime. For a de Sitter spacetime we derive the boundary condition that must be satisfied to be able to fulfill the return journey. Concluding remarks are driven in Sec.~\ref{sec5}.

\section{Mathematical background}
\label{sec2}

We begin by introducing the line element of the FLRW spacetime, which describes the metric of an expanding, homogeneous and isotropic universe
\begin{equation}\label{1}
ds^2=-c^2 dt^2+R^2(t)[d\chi^2+S_{k}^2(\chi)(d\theta^2+\sin^2\theta d\phi^2)] ,
\end{equation}
where $c$ is light speed (hereafter $c=1$),
  \[
    S_{k}(\chi)=\left\{
                \begin{array}{ll}
                  \sin\chi,&\qquad k=+1,\qquad \text{closed},\\
                  \chi,&\qquad k=0,\qquad \quad \text{flat} ,\\
                  \sinh\chi,&\qquad k=-1,\qquad \text{open},
                \end{array}
              \right.
  \]
expresses the space curvature and $R(t)$ is the scale factor which describes the expansion of the universe. $t$ is the coordinate time $t\in [0,\infty )$; $\chi$ lies in
the range $\chi \in [0,\infty )$ for $k=0,-1$ and $\chi \in [0,\pi ]$ for $k=1$; while the angles $\theta \in [0,\pi ]$ and $\phi \in [0,2\pi )$ independently of the curvature.

Let us assume a cosmological model with a cosmological constant $\Lambda$ and a fluid with equation of state given by
\begin{equation}\label{eq:EqOfState}
    P=P(\rho)=(\gamma-1)\rho\, 
\end{equation}
where $P$ is the pressure, $\rho$ is the energy density and we assume that constant $\gamma$ can take any values. Then, the Friedmann equation reads
\begin{equation}\label{Friedmann}
\frac{\dot{R}^2(t)}{R^2(t)}=\frac{\Lambda}{3}-\frac{k}{R^2(t)}+\frac{C}{R^{3\gamma}(t)},
\end{equation} 
where dot means derivation with respect to $t$  and $C$ is a constant proportional to the matter density  (see e.g. \cite{griffiths2009exact}).

The four-acceleration of a particle is given by 
\begin{equation}\label{equation of motion}
a^{\mu}=u^{\mu}_{;\nu}u^{\nu}=\frac{du^{\mu}}{d\lambda}+\Gamma^{\mu}_{\nu\sigma}\frac{dx^{\nu}}{d\lambda}\frac{dx^{\sigma}}{d\lambda} ,
\end{equation}
where $u^{\mu}$ is the four-velocity and $\lambda$ is the proper time. $a^{\mu}$ and $u^{\mu}$  satisfy the following constraints
\begin{equation}\label{uu}
u^{\mu}u_{\mu}=-1,
\end{equation}
\begin{equation}\label{aa}
 a^{\mu}a_{\mu}=A^2,
\end{equation}
\begin{equation}\label{au}
a^{\mu}u_{\mu}=0,
\end{equation}
where $A$ is the norm of the acceleration. Having uniform acceleration means that $A=$const.

 Solving Eq.~\eqref{equation of motion} for a given acceleration (say for $A=$const) is almost analytically intractable (see e.g \cite{rindler1960hyperbolic}). Here we introduce the conformal time transformation in order to tackle this problem. In particular, the conformal time $\eta$ is such that
\begin{equation}\label{conformal transformation}
\eta=\int\frac{dt}{R(t)}.
\end{equation}
Additionally, by putting $\tilde{\chi}=\chi$, the FLRW metric reads 
\begin{equation}\label{conformal metric}
ds^2=\tilde{R}^2(\eta)[-d\eta^2+d\tilde{\chi}^2+S_{k}^2(\tilde{\chi})(d\theta^2+\sin^2\theta d\phi^2)],
\end{equation}
Notice that, $\tilde{R}(\eta)=R(t)$.

When we have cosmological models with $\Lambda=0$ or $\Lambda\neq 0$ but without matter, it holds that  (see e.g. \cite{griffiths2009exact})
  \[
    \tilde{R}(\eta)=\left\{
                \begin{array}{ll}
                  \tilde{R}_c\sin^b(\frac{\eta}{b}),&\qquad k=+1,\\
                  \tilde{R}_c\eta^b,&\qquad k=0,\\
                  \tilde{R}_c\sinh^b(\frac{\eta}{b}), &\qquad k=-1,
                \end{array}
              \right.
  \]
where $\tilde{R}_c$ is a constant length which determines the scale of the universe. The power coefficient $b$ for $\Lambda=0$ is $b=\frac{2}{3\gamma-2}$.  The value of $b$  distinguishes between different cosmological models. For example, if $b=2$ then the universe is filled with dust;for stiff matter $b=\frac{1}{2}$; while $b=1$ describes the radiation case. Moreover, for non-negative curvature when $\Lambda\neq 0$ and without matter, which is actually a de-Sitter cosmological model, then $b=-1$. 

\section{Path of particles in FLRW spacetime}
\label{sec3}
We would like first to present the general formulation for the motion of particles in the transformed FLRW spacetime \eqref{conformal metric} by considering only the radial motion. To do that, we shall define the four-velocity as follows \cite{heyl2005long} 
\begin{equation}\label{velocity}
u^{\eta}=\frac{d\eta}{d\lambda}=\frac{\cosh\zeta(\lambda)}{\tilde{R}(\eta)},\quad u^{\tilde{\chi}}=\frac{d\tilde{\chi}}{d\lambda}=\frac{\sinh \zeta(\lambda)}{\tilde{R}(\eta)} ,
\end{equation}
where $\zeta(\lambda)$ is the rapidity, which will be determined later. Note that equations~\eqref{velocity} automatically satisfy constraint~\eqref{uu}. 

 The only needed non-vanishing Christoffel symbols for this case are 
\begin{equation}\label{gamma}
\Gamma^{\eta}_{\eta\eta}=\Gamma^{\eta}_{\tilde{\chi}\tilde{\chi}}=\Gamma^{\tilde{\chi}}_{\eta\tilde{\chi}}=\frac{1}{\tilde{R}(\eta)}\frac{d \tilde{R}(\eta)}{d\eta}.
\end{equation} 

The four-acceleration in the set of coordinates \eqref{conformal metric} can be written in the following way
\begin{equation}\label{eqat}
a^{\eta}=\frac{d u^{\eta}}{d\lambda}+\Gamma^{\eta}_{\eta\eta} (u^{\eta})^2+\Gamma^{\eta}_{\tilde{\chi}\tilde{\chi}}(u^{\tilde{\chi}})^2,
\end{equation}
\begin{equation}
a^{\tilde{\chi}}=\frac{du^{\tilde{\chi}}}{d\lambda}+2\Gamma^{\tilde{\chi}}_{\eta\tilde{\chi}}u^{\eta}u^{\tilde{\chi}}.
\end{equation}

From now on, since all used Christoffel symbols have equal value, we shall denote them by $\Gamma$.

By differentiating the first term in the right-hand side of Eq.~\eqref{eqat} and by using Eq.~\eqref{velocity} we obtain
\begin{equation}\label{utl} 
\frac{d u^{\eta}}{d\lambda}=\frac{\sinh\zeta(\lambda)}{\tilde{R}(\eta)}\frac{d\zeta(\lambda)}{d\lambda}-\frac{\cosh\zeta(\lambda)}{\tilde{R}(\eta)^2}\frac{d \tilde{R}(\eta)}{d\lambda}.
\end{equation}
Since the $\frac{d \tilde{R}(\eta)}{d\lambda}$ can be written in terms of $\eta$ 
\begin{equation}
\frac{d \tilde{R}(\eta)}{d\lambda}=\frac{d\eta}{d\lambda}\frac{d \tilde{R}(\eta)}{d \eta}=u^{\eta}\frac{d\tilde{R}(\eta)}{d\eta},
\end{equation}
by substituting it into Eq.~\eqref{utl} together with Eqs.~\eqref{velocity} and \eqref{gamma} we arrive to
\begin{equation}
\frac{d u^{\eta}}{d\lambda}=u^{\tilde{\chi}}\frac{d\zeta(\lambda)}{d\lambda}-(u^{\eta})^2\Gamma.
\end{equation}
Thus 
\begin{equation}
a^{\eta}=u^{\tilde{\chi}}\frac{d\zeta(\lambda)}{d\lambda}+(u^{\tilde{\chi}})^2\Gamma=u^{\tilde{\chi}}(\frac{d\zeta(\lambda)}{d\lambda}+\Gamma u^{\tilde{\chi}}).
\end{equation}

Similar calculation can be undertaken for $a^{\tilde{\chi}}$ where
\begin{equation}
\frac{du^{\tilde{\chi}}}{d\lambda}=u^{\eta}\frac{d\zeta(\lambda)}{d\lambda}-u^{\eta}u^{\tilde{\chi}}\Gamma,
\end{equation}
which finally gives
\begin{equation}
a^{\tilde{\chi}}=u^{\eta}\frac{d\zeta(\lambda)}{d\lambda}+u^{\eta}u^{\tilde{\chi}}\Gamma=u^{\eta}(\frac{d\zeta(\lambda)}{d\lambda}+\Gamma u^{\tilde{\chi}}).
\end{equation}

We denote 
\begin{equation}\label{Aformula}
A=\frac{d\zeta(\lambda)}{d\lambda}+\Gamma u^{\tilde{\chi}},
\end{equation}
where $A$, is the norm of acceleration as mentioned earlier in Eq.~\eqref{aa}. As a result, the four-acceleration becomes
\begin{equation}\label{foura}
a^{\eta}=A u^{\tilde{\chi}}, \qquad a^{\tilde{\chi}}=A u^{\eta}.
\end{equation}
Note that Eq.~\eqref{foura} satisfies also the constraints \eqref{aa} and \eqref{au}.

In the transformed FLRW metric~\eqref{conformal metric} the coordinates $\eta$ and $\tilde{\chi}$ share a common coefficient, i.e. the $\tilde{R(\eta)}$. If we constraint the motion only on the radial direction through this transformation we get a solvable set of equations from Eq.~\eqref{equation of motion}. This allows us to analyze the radial motion of the rocketeer in the FLRW spacetime.
 
\subsection{Accelerated Radial Motion}
It is convenient to express the equation of motion of the rocketeer in terms of $\eta$. Therefore, from the four-velocity \eqref{velocity}, we get
\begin{equation}
\frac{d\tilde{\chi}}{d \eta}= \frac{d\tilde{\chi}/ d\lambda}{d\eta /d\lambda}=\tanh\zeta(\lambda).
\end{equation}
To find the  unknown rapidity $\zeta(\lambda)$ we need to use Eq.~\eqref{Aformula} and reparametrize it in terms of $\eta$
\begin{equation}\label{AA}
A-u^{\eta}\frac{d\tilde{\zeta}(\eta)}{d\eta}-\Gamma u^{\tilde{\chi}}=0,
\end{equation}
where $\tilde{\zeta}(\eta)=\tilde{\zeta}(\eta(\lambda))=\zeta(\lambda)$ and $A=$const..

 Integrating Eq.~\eqref{AA} with respect to $\eta$, we obtain
\begin{equation}
\tilde{\zeta}(\eta)=\arcsinh(A\tilde{\mathscr{R}}(\eta)+\frac{\upsilon}{\tilde{R}(\eta)}),\label{zeta}
\end{equation}
where
\begin{equation}\label{eq:intr.conf}
\tilde{\mathscr{R}}(\eta)=\frac{\int^{\eta} \tilde{R}(\tilde{\eta})^2 d\tilde{\eta}}{\tilde{R}(\eta)},
\end{equation}
and $\upsilon$ is an integration constant which is related to the initial velocity of particle. Consequently,
\begin{multline}
\tilde{\chi}=A\int \frac{\tilde{\mathscr{R}}(\eta)}{\sqrt{(A\tilde{\mathscr{R}}(\eta)+\frac{\upsilon}{\tilde{R}(\eta)})^2+1}} d\eta
\\
+ \upsilon\int \frac{1/\tilde{R}(\eta)}{\sqrt{(A\tilde{\mathscr{R}}(\eta)+\frac{\upsilon}{\tilde{R}(\eta)})^2+1}} d\eta. \label{confac}
\end{multline}

Now, we go back to the coordinates of the original FLRW metric~\eqref{1}. This is achieved by using the inverse transformation of Eq.~\eqref{conformal transformation}, i.e. $\tilde{R}(\eta) d \eta=dt$, and by recalling that $R(t)=\tilde{R}(\eta)$. Thus, we obtain for Eq.~\eqref{zeta} 
\begin{equation}
    \hat{\zeta}(t)=\arcsinh(A\mathscr{R}(t)+\frac{\upsilon}{R(t)}),\label{zetaflrw}
\end{equation}
where
\begin{equation}\label{eq:intr}
\mathscr{R}(t)=\frac{\int^t R(t^{\prime}) dt^{\prime}}{R(t)}.
\end{equation}

Using Eq.~\eqref{zetaflrw} enables us to derive the four-velocity in standard FLRW spacetime as follows
\begin{equation}\label{standardvelocity}
u^{t}=\frac{d t}{d\lambda}=\cosh\hat{\zeta}(t),\quad u^{\chi}=\frac{d\chi}{d\lambda}=\frac{\sinh\hat{\zeta}(t)}{R(t)}.
\end{equation}

Finally, the trajectory of uniform acceleration motion is given by 
\begin{multline}
\chi=A\int \frac{1}{R(t)}\frac{\mathscr{R}(t)}{\sqrt{(A\mathscr{R}(t)+\frac{\upsilon}{ R(t)})^2+1}} dt
\\
+\upsilon\int \frac{1}{R(t)^2}\frac{1}{\sqrt{(A\mathscr{R}(t)+\frac{\upsilon}{ R(t)})^2+1}} dt,\label{accel}
\end{multline}

By specifying the evolution of the scale factor $R(t)$ Eq.~\eqref{accel} provides the accelerated radial path of the rocketeer in the standard FLRW coordinate.

\subsection{Some characteristic types of motion}\label{subsec:feature}
\paragraph{Purely accelerated motion.} When  
one ignores the integration constant $\upsilon$ (i.e. $\upsilon=0$) in trajectories~\eqref{confac} and~\eqref{accel} the motion is called \textit{purely accelerated}. In the conformally transformed coordinates the trajectory is given by
\begin{equation}\label{pconac}
    \tilde{\chi}_a=A\int \frac{\tilde{\mathscr{R}}(\eta)}{\sqrt{A^2\tilde{\mathscr{R}}(\eta)^2+1}} d\eta,
\end{equation}
and in the original FLRW coordinates we get 
\begin{equation}\label{paccel}
    \chi_a=A\int \frac{1}{R(t)}\frac{\mathscr{R}(t)}{\sqrt{A^2\mathscr{R}(t)^2+1}} dt,
\end{equation}
where index $a$ in both Eqs.~\eqref{pconac} and \eqref{paccel} refers to the \textit{purely accelerated motion}.

\paragraph{Geodesic motion.}
To get the trajectory for the geodesic motion one has to substitute $A=0$ into the Eqs.~\eqref{zeta}, \eqref{confac} and \eqref{accel}. Thus, the rapidity function $\tilde{\zeta}(\eta)$  becomes
\begin{equation}
\tilde{\zeta}(\eta)=\arcsinh\Big(\frac{\upsilon}{ \tilde{R}(\eta)}\Big).
\end{equation}

Consequently, Eq.~\eqref{confac} will be 
\begin{equation}\label{confv}
\tilde{\chi}_v=\int \frac{\upsilon}{\sqrt{ \tilde{R}(\eta)^2+\upsilon^2}} d\eta,
\end{equation}
and for Eq.~\eqref{accel} we obtain
\begin{equation}\label{veloc}
\chi_v=\int \frac{\upsilon}{R(t)}\frac{1}{\sqrt{ R(t)^2+\upsilon^2}} dt.
\end{equation}
Here index $v$ in Eqs.~\eqref{confv} and \eqref{veloc} denotes geodesic motion. For all $\upsilon$ values, $\upsilon^2>0$, which guarantees that Eq.~\eqref{veloc} is a timelike geodesic \cite{cunningham2017exact}.

Eqs.~\eqref{confv} and \eqref{veloc} are geodesics in any conformal time FLRW spacetime and FLRW spacetime respectively. Eq.~\eqref{veloc} is the same as equation derived in \cite{gron2007space} and recently in \cite{cunningham2017exact}.
\paragraph{Null geodesics.}
 We can see from Eq.~\eqref{accel} that for large acceleration $A$ the particle's trajectory asymptotically reaches the null geodesic, that means 
 \begin{equation}
\lim_{A>>} \chi_a=\pm\int \frac{1}{R(t)}dt.
 \end{equation}
 Moreover, this statement holds for large $\upsilon$ value, i.e.
 \begin{equation}
\lim_{\upsilon>>} \chi_a=\pm\int \frac{1}{R(t)}dt.
 \end{equation}
 \paragraph{Transformation conditions.} For the flat spatial curvature FLRW spacetime the accelerated motion~\eqref{pconac}  can be transformed into geodesic motion~\eqref{confv} under certain conditions. In order to investigate this statement, we consider two different spacetimes having scale factors $\tilde{R}_m(\eta)= \mu \eta^m$ and $\tilde{R}_k(\eta)= \kappa \eta^k$. By substituting $\tilde{R}_m(\eta)$ and $\tilde{R}_k(\eta)$ into the Eqs.~\eqref{confv} and~\eqref{pconac} respectively we get 
 \begin{equation}
    \tilde{\chi}_v=\frac{\upsilon\,\eta^{1-m}}{\mu\,(m-1)}{\mbox{$_2$F$_1$}}\left(\frac{1}{2},\frac{m-1}{2\,m};\frac{3\,m-1}{2\,m};-\frac{\upsilon^2\,\eta^{-2\,m}}{\mu^2} \right),
 \end{equation}
 and
 \begin{equation}
     \tilde{\chi}_a=\frac{A\,\kappa\,\eta^{2+k}}{(2+k)\,(1+2\,k)}{\mbox{$_2$F$_1$}}\left(\frac{1}{2},\frac{2+k}{2+2\,k};\frac{4+3\,k}{2+2\,k};-\frac{A^2\,\kappa^2\,\eta^{2+2k}}{(1+2\,k)^2} \right),
 \end{equation}
where ${\mbox{$_2$F$_1$}(a,b;\,c;\,z)}$ is the Gauss hypergeometric function. These two trajectories become equivalent if 
 \begin{equation}\label{condition}
    \left\{
    \begin{array}{c}
     1+k=-m,\\
     \frac{A\,\kappa}{1+2\,k}=\frac{\upsilon}{\mu},\\
     1+2\,k\neq0.
     \end{array}
     \right.
    \end{equation}
 For instance, the uniformly accelerated trajectories in the de Sitter spacetime get transformed to the geodesic motion in the Minkowski spacetime and vice versa ( See Sec.~\ref{subsec:dess}  for more details). Moreover, one can show that an observer with a suitable acceleration moving in a decelerating Friedmann universe, i.e. a dust field universe, has the same cosmological redshift as the observer in the $\Lambda$CDM model~\cite{feoli2017dark}.
 
\section{Some Examples }\label{sec4}
In this section of section's \ref{sec3} formalism is applied to specific FLRW universe models. The motion of a particle both in the original and in the transformed coordinates depends only on the scale factor and its integral (Eq.~\eqref{eq:intr} and Eq.~\eqref{eq:intr.conf} respectively). Thus, specifying the expansion factor for each cosmological model enables us to determine the particles worldlines. In this section the behavior of the trajectories presented in paragraphs a and b of Sec.~\ref{subsec:feature} is studied.

Recently, the solution of Friedmann Eq.~\eqref{Friedmann} was presented for various FLRW models with $k=0$  \cite{chavanis2015cosmology}. Namely, Chavanis has derived an analytical solution for $R(t)$ in a universe undergoing a various combination of eras, e.g. stiff matter era, dark matter era, and dark energy era due to the cosmological constant. From this study we use the form of the scale factor in the cosmological examples of Sec.~\ref{sec4} and Sec.~\ref{sec:return}.

Note that, although the transformation \eqref{conformal transformation} is not, in general, conformally flat transformation (CFT) for spatially curved FLRW models, it is CFT for all the flat FLRW models. Thus, since the cosmological models appearing in  Secs.~\ref{generalscalefactor} and~\ref{subsec:dess} have zero spatial curvature,  the transformation \eqref{conformal transformation} is a CFT, i.e. it holds that
\begin{equation}
ds^2_{k=0}=\Omega^2 ds^2_{flat},
\end{equation} 
where $\Omega=\tilde{R}(\eta)$.

It is clear that  this  formalism is able to reproduce the known hyperbolic motion in the Minkowski spacetime~\cite{rindler2012essential}. In a similar manner as in the Minkowski spacetime, the uniformly accelerated motion can be derived in the Einstein static universe, since for both spacetimes the scale factor $R(t)=1$.

To provide visualization for our examples we are going to plot some trajectories in Penrose diagrams with coordinates $\eta$ and $\chi$ given by the metric~\eqref{conformal metric}.

\subsection{Flat FLRW spacetime without $\Lambda$}\label{generalscalefactor}
In this section we consider spatially flat FLRW spacetime with vanishing cosmological constant, i.e. $\Lambda=0$, and a single fluid content provided by the EoS~\eqref{eq:EqOfState}. From the Friedmann equation~\eqref{Friedmann} one can obtain the scale factor 
\begin{equation}
    R(t)= R_c t^{\frac{2}{3\,\gamma}},
\end{equation}
where $R_c=\left(\frac{3}{2}\,\gamma\,\sqrt{C}\right)^{\frac{2}{3\,\gamma}}$. Substituting this scale facor into  the equations Eqs.~\eqref{paccel} and \eqref{veloc} we obtain
\begin{multline}
\chi_a=\frac{9\,A\,t^{2-2/3\,\gamma}\,\gamma^2}{R_c\,(3\,\gamma+2)\,(6\,\gamma-2)}\\
    {\mbox{$_2$F$_1$}\left(\frac{1}{2},1-\frac{1}{3\gamma};2-\frac{1}{3\gamma};-\left(\frac{3\,A\,\gamma\,t}{2+3\,\gamma}\right)^2 \right)},
\end{multline}
and 
\begin{multline}
    \chi_v=\frac{3\,t^{1-2/3\,\gamma}}{R_c\,(3-2\,\gamma)}\\
    {\mbox{$_2$F$_1$}\left(\frac{1}{2},\frac{3\,\gamma}{4}-\frac{1}{2};\frac{3\,\gamma}{4}+\frac{1}{2};-\left(\frac{R_c\,t^2/3\,\gamma}{\upsilon} \right)^2 \right)},
\end{multline}
where $a$ and $v$ denote uniform acceleration and geodesic motion respectively.
In conformal representation, where $\tilde{R}(\eta)=\tilde{R}_c \eta^b$ where $b=\frac{2}{3\,\gamma-2}$, we have
\begin{multline}
    \tilde{\chi}_a=\frac{A\,\tilde{R}_c\,\eta^{2+b}}{(2+b)\,(1+2\,b)}\\
    {\mbox{$_2$F$_1$}}\left(\frac{1}{2},\frac{2+b}{2+2\,b};\frac{4+3\,b}{2+2\,b};-\frac{A^2\,\tilde{R}_c^2\,\eta^{2+2b}}{(1+2\,b)^2} \right)
\end{multline}
and
\begin{multline}
    \tilde{\chi}_v=\frac{\upsilon\,\eta^{1-b}}{\tilde{R}_c\,(b-1)}
    {\mbox{$_2$F$_1$}}\left(\frac{1}{2},\frac{b-1}{2\,b};\frac{3\,b-1}{2\,b};-\frac{\upsilon^2\,\eta^{-2\,b}}{\tilde{R}_c^2} \right).
\end{multline}

\begin{figure}[ht]
    \centering
       {\includegraphics[width=0.45\textwidth]{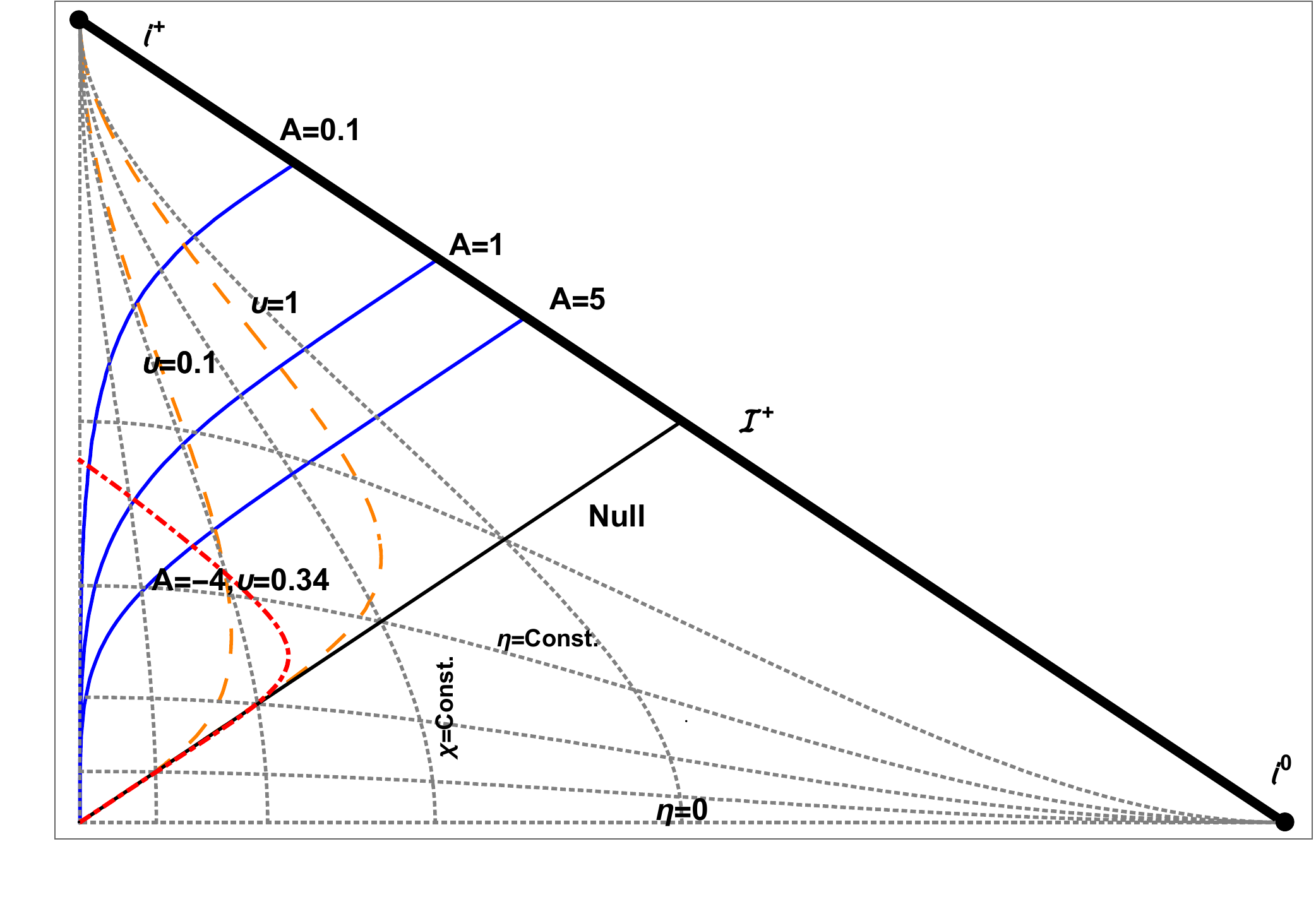}}
       \caption{Penrose diagram for the dust-filled universe. The thick black solid line denotes the photon trajectory. Solid lines have constant acceleration $A$ and $\upsilon=0$, while the dotted dashed lines have constant acceleration $A$ and $\upsilon \neq 0$. Dashed curves are geodesics, i.e. $A=0$.}\label{fig1}
\end{figure}

The dynamical features of the above trajectories on a Penrose diagram are very similar for all the usual barotropic fluids, i.e. for fluids with $1\leq \gamma\leq 2$. Thus, in Fig.~\ref{fig1} we plot just the case $\gamma=1$, which shows different trajectories in a spatially flat FLRW universe with dust. Namely, Fig.~\ref{fig1} shows accelerated trajectories with zero $\upsilon$ and non zero $\upsilon$ (solid blue and dotted dashed red lines respectively), the geodesic trajectories (orange dashed lines) together with the null geodesic (black thick solid line). In the case of constant acceleration, the greater the acceleration the faster the rocketeer approaches the null geodesic behavior. We plot also one decelerating trajectory with non-zero $\upsilon$, which exhibits a return journey: such trajectories will be discussed in Sec.~\ref{sec:return}. Regarding the geodesic motion the greater the initial velocity, the further the traveller can reach. Note that even if initially geodesic travellers overtake accelerated ones, eventually as expected the accelerated ones prevail. Additionally,  this figure provides the asymptotic behavior of the trajectories as $t\rightarrow \infty$. One can see that, accelerated trajectories reach the future null infinity ,i.e. $ \mathcal{I}^+$, whereas the geodesics motion ends up to timelike infinity, i.e. ${i}^+$.

\subsection{Milne Universe}\label{subsec:milne}
Vacuum FLRW model with $\Lambda=0$ and $k=-1$ is known as Milne universe \cite{milne1932world}, where $R(t)=t$. For this spacetime particle's paths are 
\begin{equation}
\chi_a=\ln  \left( At+\sqrt{A^2t^2+4} \right) +\ln  \left( C_5 \right),
\end{equation}
where $C_5$ is an integration constant. For a particle starting from $\chi_a=t=0$ $(C_5=\frac{1}{2})$, $\chi_a$ reduces to 
\begin{equation}\label{amil}
\chi_a=\arcsinh(\frac{A}{2}t),
\end{equation}
and $\chi_v$ becomes
\begin{equation}\label{vmil}
\chi_v=-\arctanh(\frac{\upsilon}{\sqrt{t^2+\upsilon^2}})+\arctanh(-\frac{\upsilon}{\sqrt{1+\upsilon^2}}).
\end{equation}

For Milne universe we cannot use Eqs.~\eqref{confac} and \eqref{confv} since transformation \eqref{conformal transformation} is not CFT. In order to plot the above case in a Penrose diagram we have to use the transformation 
\begin{equation}
t=\sqrt{T^2-R^2},\qquad \chi=\arctanh(\frac{R}{T}),
\end{equation}
between a Milne Universe and the Minkowski spacetime \cite{rindler2012essential}.
Using this transformation we get 
\begin{equation}\label{amink}
(R_a+\frac{1}{A})^2-T^2=\frac{1}{A^2},
\end{equation}
and
\begin{equation}\label{vmink}
R_v=\frac{\upsilon\, T}{\sqrt{1+\upsilon^2}},
\end{equation}
for the trajectories \eqref{amil} and \eqref{vmil} respectively in the Minkowski spacetime.

\begin{figure}[ht]
    \centering
       {\includegraphics[width=0.45\textwidth]{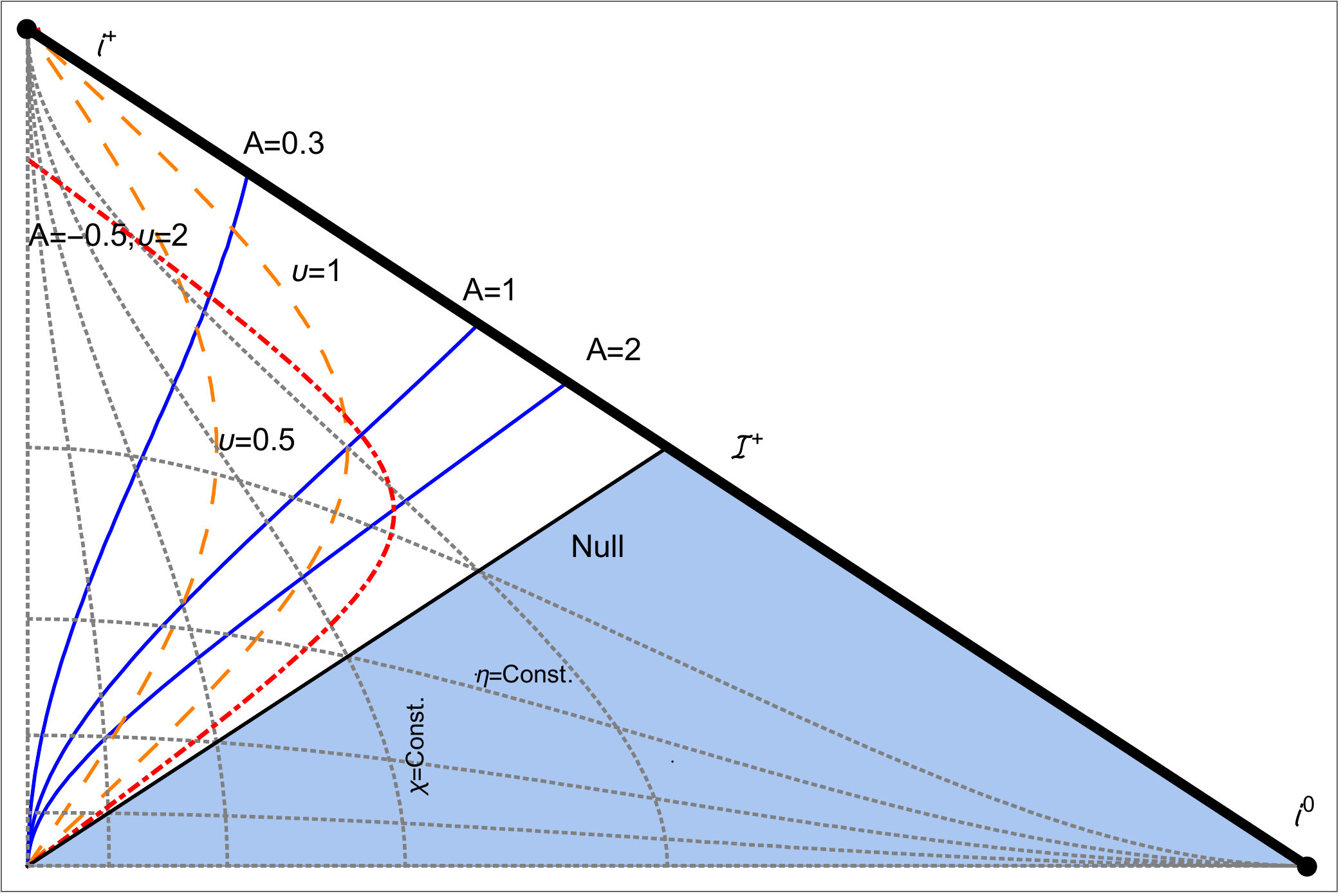}}
       \caption{Penrose diagram for the Milne universe. Curves have the same coloring as described in Fig.~\ref{fig1}. Note that, the shaded part of the plot does not belong to the Milne universe. }\label{figmilne}
\end{figure}

 It is known that Eqs.~\eqref{amink} and \eqref{vmink} describe hyperbolic and geodesic motion respectively in Minkowski spacetime. In Fig.~\ref{figmilne} examples of these types of motion are depicted in the same manner as Fig.~\ref{fig1}. The shaded region in this figure indicates the part of the Penrose diagram that does not belong to the Milne universe. 
 
\subsection{de-Sitter Universe, $\gamma=0$}\label{subsec:dess}

Considering the dark energy dominated universe in the absence of any matter the scale factor is
\begin{equation}\label{desfactor}
R(t)=R_0 e^{\sqrt{\frac{\Lambda}{3}} t},
\end{equation}
where $\Lambda$ is a cosmological constant \cite{chavanis2015cosmology}. This solution is known as the de Sitter solution.

Accelerated and geodesic motion in this particular spacetime are described by
\begin{equation}
\chi_a= -\frac{A}{\sqrt{\frac{\Lambda}{3}} R_0 e^{\sqrt{\frac{\Lambda}{3}} t}\sqrt{A^2+\frac{\Lambda}{3}}}+C_1,
\end{equation}
\begin{equation}
\chi_v=-\sqrt{\frac{3}{\Lambda}}\frac{\sqrt{(R_0 e^{\sqrt{\frac{\Lambda}{3}} t})^2+\upsilon^2 }}{\upsilon R_0 e^{\sqrt{\frac{\Lambda}{3}} t}}.
\end{equation}
In conformal coordinates, where scale factor is $\tilde{R}(\eta)=-\sqrt{\frac{3}{\Lambda}}\frac{1}{\eta}$, we get
\begin{equation}\label{dsa}
\tilde{\chi}_a=\frac{ A}{\sqrt{A^2+\frac{\Lambda}{3}}}\eta,
\end{equation}
\begin{equation}\label{dsv}
\tilde{\chi}_v=-\sqrt{\frac{3}{\Lambda\upsilon^2}+\eta^2}+C_3.
\end{equation}

\begin{figure*}[ht]
    \centering
       {\includegraphics[width=0.7\textwidth]{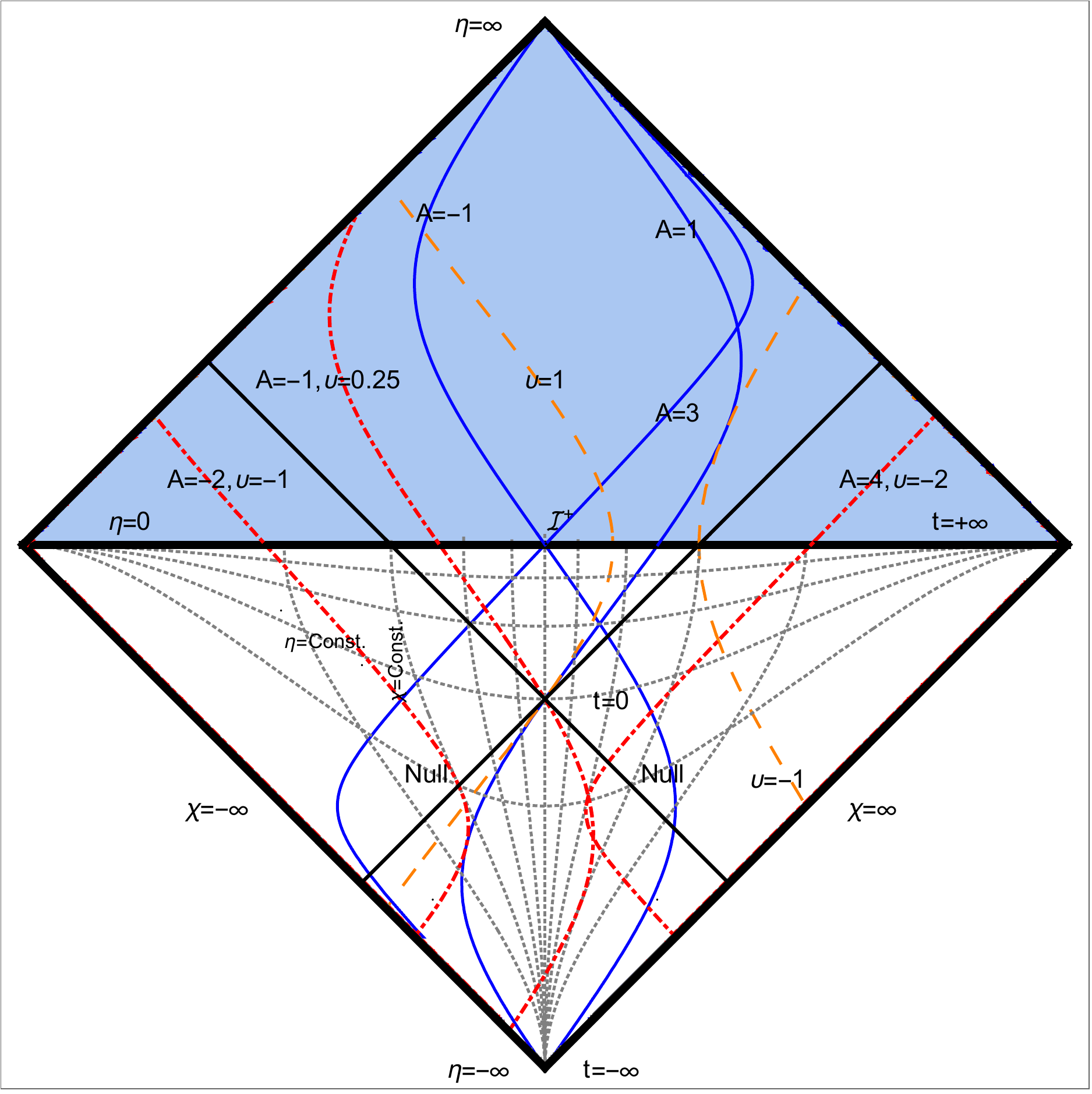}}
       \caption{Penrose diagram for de Sitter universe.The shaded part is not covered by the flat de Sitter universe. }\label{fig2}
\end{figure*}

It is clear from Eqs.~\eqref{dsa} and \eqref{dsv} that the geodesic equation in conformally flat coordinates ( Minkowski spacetime) get transformed to uniformly accelerated worldline in de Sitter spacetime, whereas the trajectory of a uniformly accelerated particle in Minkowski spacetime get transformed to geodesic in de Sitter spacetime (see Fig.~\ref{fig2}). This result confirms previous works of Rindler\footnote{Note that, W.Rindler obtained only one special case of accelerated motion in de Sitter spacetime. Namely, he studied the case when a particle leaves the origin ($t=\chi=0$) from rest, i.e. $u^t=1$ and $u^{\chi}=0$. One can rederive Rindelr's trajectory by putting $\upsilon=-\sqrt{\frac{3}{\Lambda}}A$ into the Eq.~\eqref{accel} together with $\chi_{(t=0)}=0$.}~\cite{rindler1960hyperbolic} and Bi{\v{c}}{\'a}k \&  Krtou{\v{s}}~\cite{bivcak2001j}.

Fig.~\ref{fig2} shows the trajectories in the de Sitter spacetime. In this spacetime, all trajectories have the same description as in Fig~\ref{fig1}, but some of them have different initial conditions. Namely, some of them do not pass through the origin $t=0$. Another difference is that the de Sitter spacetime covers only the lower part of the Penrose diagram, since all trajectories end up at the $\mathcal{I}^+$. We continue  plotting the trajectories even to the upper shaded region in order to provide a global view of the behavior of these trajectories. 

\subsection{Anti-de Sitter spacetime}\label{sec:antidess}
In this section we consider a vacuum FLRW universe with a negative cosmological constant $\Lambda$ and negative spatial curvature $k=-1$ namely Anti-de Sitter universe. This particular case of Anti-de Sitter universe has the scale factor
\begin{equation}
    R(t)= \alpha\, \cos(\frac{t}{\alpha})
\end{equation}
where $\displaystyle \alpha=\sqrt{3/\mid \Lambda\mid}$. Thus, the accelerated and geodesic trajectories become
\begin{equation}
    \chi_a=\ln\left(2\frac{A^2\,\alpha^2+A\,\alpha\,\sqrt{A^2\,\alpha^2\,\sin^2(t/\alpha)+\cos^2(t/\alpha)}}{\cos(t/\alpha)} \right),
\end{equation}
and
\begin{equation}
    \chi_v=\arctanh\left(\frac{\upsilon \sin(t/\alpha)}{\sqrt{\alpha^2\,\cos^2(t/\alpha)+\upsilon^2}} \right).
\end{equation}
 \begin{figure}[ht]
    \centering
       {\includegraphics[width=0.45\textwidth]{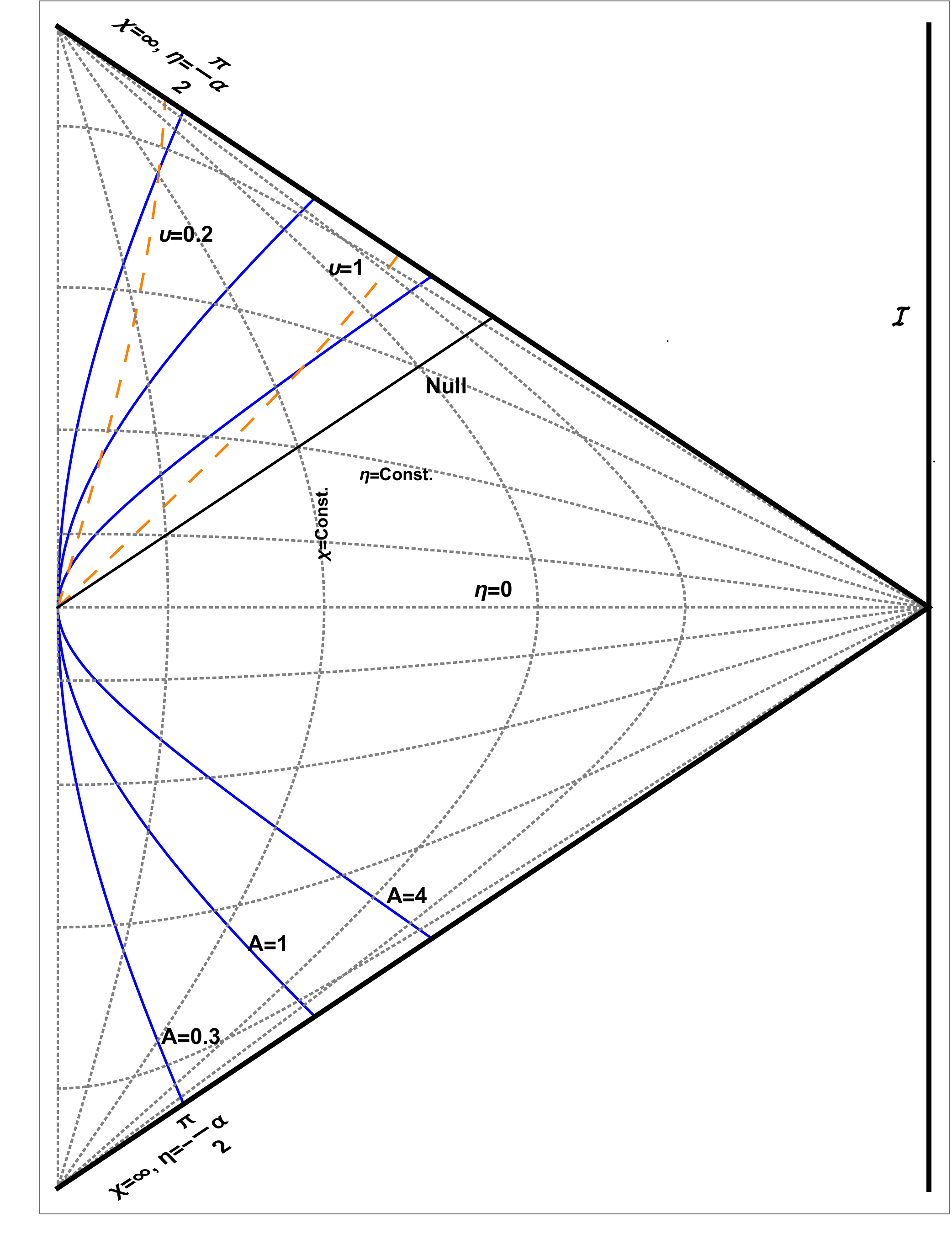}}
       \caption{Penrose diagram of the part of the anti-de Sitter spacetime as a particular case of the FLRW universe. Trajectories are collored in the same manner as Fig.~\ref{fig1}.}\label{fig:AntidessFLRW}
\end{figure}
In Fiq.~\ref{fig:AntidessFLRW} we illustrate these trajectories and we denote the different types of trajectories as we did in Fig.~\ref{fig1}.

Note that this coordinate system does not cover the whole anti-de Sitter spacetime. In order to study the accelerated motion in the whole anti-de Sitter spacetime we use the following line element
\begin{align}\label{eq:antidessmetricc}
    ds^2=- \cosh^2(r) dt^2 + \alpha^2\,\left( dr^2+\sinh^2(r)\,(d\theta^2+\sin^2\theta\,d\phi^2)\right),
\end{align}
where $r$ is dimentionless. The accelerated and geodesic motions of this metric can not be studied from the formulation presented in Sec.~\ref{sec3}. However, we can introduce a similar prescription to obtain those trajectories. Namely, we introduce the conformal coordinate $\chi$ by setting $\sinh(r)=\tan(\chi)$ together with $t=\alpha\,\eta$ . Then, the metric~\eqref{eq:antidessmetricc} takes the form 
\begin{equation}\label{eq:anticonfmetric}
    ds^2=\frac{\alpha^2}{\cos^2(\chi)}[-d\eta^2+d\chi^2+\sin^2(\chi)(d\theta^2+\sin^2\theta\,d\phi^2)].
\end{equation}
It is clear that the anti-de Sitter spacetime time covers only half of the Einstein static universe, namely in the range $\chi \in [0,\pi/2)$.

Now, similarly to the formulation presented in Sec.~\ref{sec3}, we introduce the radial four-velocity 
\begin{align}\label{eq:anti4vel}
    u^\eta=\frac{d\eta}{d\lambda}=\frac{\cosh(\xi(\lambda))}{F(\chi)},\quad u^{\chi}=\frac{d\chi}{d\lambda}=\frac{\sinh(\xi(\lambda))}{F(\chi)},
\end{align}
where $\displaystyle F(\chi)=\frac{\alpha}{\cos(\chi)}$. This radial motion has the four-acceleration given by
\begin{align}
    a^\eta=A u^{\chi},\quad a^{\chi}=A u^\eta,
\end{align}
where 
\begin{equation}\label{eq:antiAconst}
    A=\frac{d\xi(\lambda)}{d\lambda}+\Gamma u^\eta
\end{equation}
 and $  \Gamma=\frac{1}{F(\chi)}\frac{dF(\chi)}{d\chi}$.
Therefore, for the rapidity function $\xi(\chi)=\xi(\chi(\lambda))=\xi(\lambda)$,  which can be determined from  Eq.~\eqref{eq:antiAconst}, we get 
\begin{equation}\label{eq:antidessrapidity}
    \xi(\chi)=\arccosh\left( A \mathscr{F}(\chi)+\frac{\upsilon}{F(\chi)}\right),
\end{equation}
where 
\begin{equation}
    \mathscr{F}(\chi)= \frac{\int F(\hat{\chi})^2 d\hat{\chi}}{F(\chi)},
\end{equation}
and $\upsilon$ is an initial velocity of the accelerated particle. Thus, from the four-velocity~\eqref{eq:anti4vel} and Eq.~\eqref{eq:antidessrapidity}  we get
\begin{equation}\label{eq:antigentraj}
  \eta=\int \coth\left(\xi(\chi) \right) d\chi.
\end{equation}
By substituting $\upsilon=0$ into the  Eq.~\eqref{eq:antigentraj}, namely for the \textit{purely accelerated motion}, after some manipulation we get the following trajectory
\begin{equation}\label{eq:pacantidess}
    \chi_a=\pi-\arccos\left(\frac{\sqrt{A^2\,\alpha^2-1}}{A\,\alpha}\,\frac{\tan(\Delta\,\eta)}{\sqrt{1+\tan^2(\Delta\,\eta)}} \right),
\end{equation}
where $A^2\,\alpha^2>1$ has to be satisfied, $\Delta\,\eta=\eta-\eta_0$ and the constant $\eta_0$  is an integration constant from integral appears in Eq.~\eqref{eq:antigentraj}. Observers with an acceleration $A$ travel radially in the range   
$$ 
\arcsin\left(\frac{1}{A\,\alpha}\right)< \chi_a< \frac{\pi}{2}.
$$
At $\displaystyle \chi_a=\arcsin\left(1/(A\,\alpha)\right)$ the $\chi$ component of the four-velocity vanishes and the radial moving observer turns to the stationary observer. Furthermore, Eq.~\eqref{eq:pacantidess} show that the maximum duration of the radial accelerated traveler in the anti-de sitter spacetime  is $\Delta\,\eta=\pi/2$, since the trajectories reach the $\mathcal{I}$.

Moreover, by putting $A=0$ into the Eq.~\eqref{eq:antigentraj} we derive the trajectories for geodesic motion
\begin{equation}\label{eq:geoantidess}
    \chi_v=\arcsin\left(\frac{\sqrt{\upsilon^2-\alpha^2}}{\upsilon}\,\frac{\tan(\Delta\,\eta)}{\sqrt{1+\tan^2(\Delta\,\eta)}} \right),
\end{equation}
which holds under the condition that $\upsilon^2>\alpha^2$ (for $\upsilon=\alpha$ the geodesic trajectory vanishes). Eq.~\eqref{eq:geoantidess} shows that, the observer moving with the constant $\upsilon$ moves in the range  $$0<\chi_v<\arccos\left(\frac{\alpha}{\upsilon}\right).$$ When the observer reaches at $\chi_v=\arccos\left(\alpha/\upsilon\right)$ the $u^\chi$ becomes zero and therefore the radial moving observer becomes stationary. Similar to the accelerated motion, maximum duration of this motion is $\pi/2$. 

Setting $\xi(\chi_0)=0$ for any fixed $\chi=\chi_0$ reduces the accelerated radial motion to the family of the  \textit{timelike worldlines} representing uniformly  accelerated observers studied in previous works \cite{griffiths2009exact,podolsky2002accelerating,krtouvs2005accelerated}. Thus, the newly found radially moving accelerated observers have as a limiting case the already known stationary ones.

\begin{figure*}[ht]
    \centering
       {\includegraphics[width=0.60\textwidth]{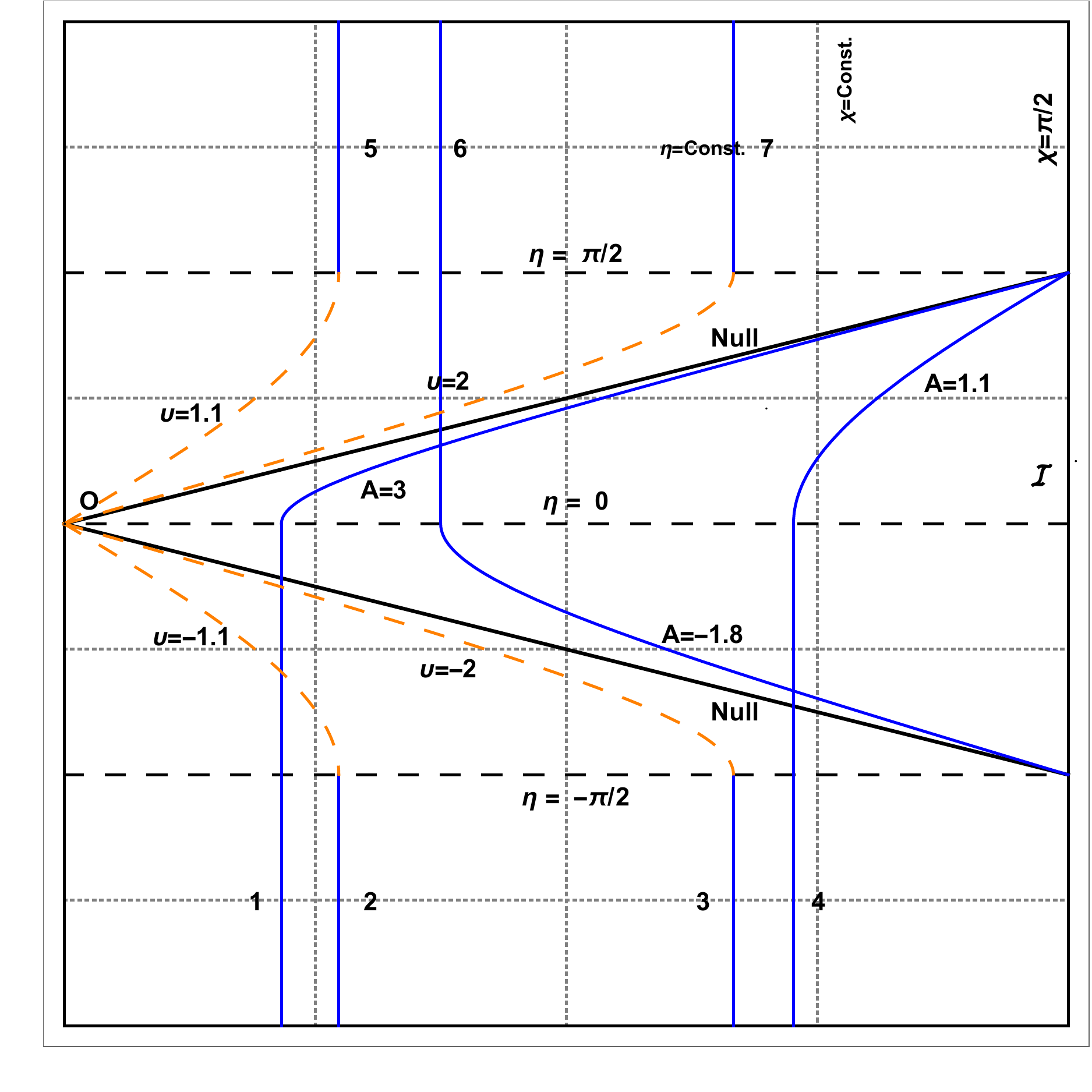}}
       \caption{The Penrose diagram for the global anti-de Sitter spacetime. See the text for more details.  }\label{fig:Antidess}
\end{figure*}

Fig.~\ref{fig:Antidess} shows these trajectories denoted in the same manner as in Fig.~\ref{fig1}. As we discussed previously, worldlines of fixed $\chi$ represent uniformly accelerated observers. In this figure, trajectory number (1) shows that the stationary observer from $\eta \in (-\infty,0]$. Then, at $\eta=0$ it starts to accelerated radially with an acceleration $A=3$ and goes toward $\mathcal{I}$( trajectory (4) has the similar behavior). Observer number (6) has a deceleration $A=-1.8$ from $\eta \in (-\pi/2,0)$. Then, at $\eta=0$ its $u^\chi$ vanishes and becomes stable. On the other hand, trajectories (5) and (7) which have the $\upsilon>0$ start from $\eta=\chi=0$ traveling with for $\Delta\,\eta=\pi/2$ with constant $\upsilon$. Its radial component of four-velocity, i.e. $u^\chi$ is decreasing until at $\eta=0$ it become zero. After this point, the observer becomes stationary. Observers (2) and (3) are at rest from $\eta \in (-\infty,-\pi/2]$ and then they move with negative $\upsilon$ towards $O$. 

\section{Return Journey }\label{sec:return}
In this section we are going to focus our study on analyzing the return journey of the rocketeer in spatially flat FLRW universe. In particular, we are going to study the behavior of  the Eq.~\eqref{accel} or Eq.~\eqref{confac} in the spacetimes studied in the previous section. Actually, in order to fulfill the return journey, our rocketeer must begin to decelerate, i.e. $A_d<0$, long enough time before it reaches the designated proper distance.

Assuming that, the spaceship is travelling with non-zero positive value $\upsilon$, at $t=\lambda=\chi=0$ the rocketeer applies a deceleration $A_d$. Thus, the rocketeer reaches the maximum comoving distance from the origin at the return point with coordinates $\{t_1,x_1\}>0$ when $  \hat{\zeta}(t_1)=0$ or equivalently $u^{\chi}(t_1)=0$.

Therefore, depending on the form of scale factor, one can analyze the return journey of the rocketeer. 
\subsection{$R(t)= t^n$ spacetimes}
In this section we analyze the return journey in the spacetime studied in sections~\ref{generalscalefactor}-~\ref{subsec:milne} , namely spacetimes having the scale factor like $R(t)= t^n$. In these particular spacetimes the rockeeter reaches the maximum comoving distance from the origin at 
\begin{equation}
 t_1=\left( -\frac{\upsilon(n+1)}{A_d}\right) ^{\frac{1}{n+1}}.
\end{equation}
Afterwards, the rocketeer returns towards the origin. As $t\rightarrow \infty$ the trajectory of the rocketeer asymptotically becomes
\begin{equation}
    \lim_{t\rightarrow \infty} \chi=-\int \frac{1}{t^n}dt,
\end{equation}{}
which means that there is a finite $t_2>t_1$, when the rocketeer arrives back to the origin, $\chi=0$. In Figs.~\eqref{fig1}-~\eqref{figmilne}, the dotted dash trajectories represent the return journeys in each spacetimes. 

\subsection{de-Sitter case}

\begin{figure}[ht]
    \centering
       {\includegraphics[width=0.45\textwidth]{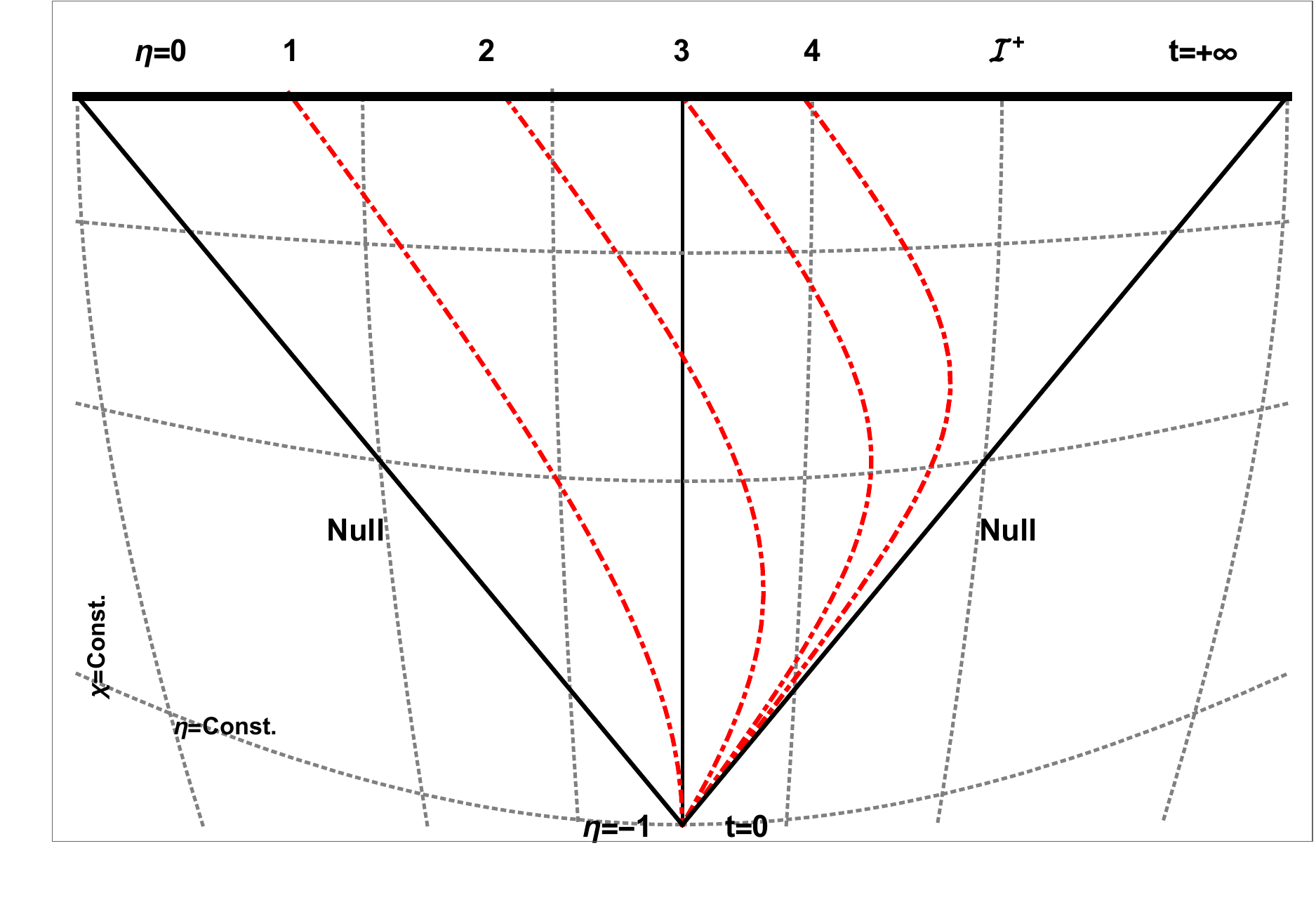}}
       \caption{Penrose diagram for the worldlines of return journeys in de Sitter spacetime. Here $\Lambda=3$ and $A=-2$ and the spacetime is depicted only for $t\geq 0$. }\label{figreturn}
\end{figure}

The return journey in the de Sitter spacetime has a different behavior with respect to the previous examples, since the scale factor~\eqref{desfactor} is given by a different function of time. In this specific spacetime, the rocketeer reaches the maximum comoving distance from the origin at 
\begin{equation}\label{toj}
    t_{1}=\sqrt{\frac{3}{\Lambda}} \ln\left(-\sqrt{\frac{\Lambda}{3}}\frac{\upsilon}{ A_d}\right).
\end{equation}
Moreover, the total cosmic time $t_2$ needed to cover the return journey for a rocketeer that leaves the origin at $t=\lambda=0$, is derived from Eq.~\eqref{accel} and it is given by
\begin{equation}\label{tret}
   t_2={\sqrt{\frac{3}{\Lambda}}\ln\left( -{\frac {\sqrt {3\Lambda}\upsilon}{\sqrt {3\Lambda}\upsilon+6 A_d}}
 \right) }.
\end{equation}
Thus, from Eqs.~\eqref{toj} and~\eqref{tret} one can see that the return journey does not happen for all values of $A_d$ and $\upsilon$ (see Fig.~\ref{figreturn}). Actually, to attain an actual return journey, the following relation
\begin{equation}\label{bound}
    2 A_d < -\sqrt{\frac{\Lambda}{3}}\upsilon< A_d,
\end{equation}
has to be satisfied. In Fig.~\ref{figreturn} we show several cases of return journeys in de Sitter spacetime for $\Lambda=3$ and $A=-2$. The negative values of $\chi$ represents the opposite direction from the one that the rocketeer is supposed to explore.
\begin{description}
    \item[Line 1.] For $-\sqrt{\frac{\Lambda}{3}}\upsilon\geq A_d$, there isn't any return point for the particle and rocketeer will move toward the $-\chi$ direction.
    \item[Line 2.] For $2 A_d < -\sqrt{\frac{\Lambda}{3}}\upsilon< A_d$, there is a return point and the rocketeer is able to come back to the origin.
    \item[Line 3.] For $-\sqrt{\frac{\Lambda}{3}}\upsilon=2 A_d$, there is a return point but the rocketeer will return back to origin in a infinite cosmic time.
    \item[Line 4.]  For $-\sqrt{\frac{\Lambda}{3}}\upsilon > 2 A_d$, there is a return point but the rocketeer will never go back to the origin.
\end{description}

Thus, we have seen that in de Sitter spacetime, having the uniform deceleration motion is not sufficient for the rocketeer to come back to the origin. One has to apply the deceleration which satisfies Eq.~\eqref{bound}.

\section{conclusions}\label{sec5}

In 1960, W.Rindler~\cite{rindler1960hyperbolic} proposed the problem of a "Hyperbolic Motion in Curved Spacetime" to study the accelerated motion in curved spacetime.

In particular, it was suggested that the accelerated motion is the best way of exploring our universe in a reasonably short time \cite{kwan2010adventures}. Taking the above suggestion into account, the motion of an accelerated traveller in an expanding universe has been studied in this work. This involves solving the non-trivial Eq.~\eqref{equation of motion} for a given acceleration. To achieve this, we applied a conformal time transformation~\eqref{conformal transformation} to the generic FLRW universe. Using the method introduced in Sec.~\ref{sec3} has helped us to determine a generalized form of rapidity function~\eqref{zetaflrw}, which leads us to derive the trajectory~\eqref{accel} of an accelerated traveler.
 
We have shown that the accelerated and the geodesic motion in an expanding universe are solely determined by the expansion factor and its integral~\eqref{eq:intr}. The scale factor is the solution of the Friedmann equation~\eqref{Friedmann}, which depends on the spatial curvature $k$, the cosmological constant $\Lambda$, and the equation of state $P=P(\rho)$. Although, we have chosen a specific form of the equation of state in Sec.~\ref{sec2}, this formulation is independent of the choice of an equation of state. It depends only on whether the scale factor can be expressed analytically as a function of time or not. 

Additionally, we have provided a similar formulation in the case of the anti-de Sitter spacetime for 
the uniformly accelerated and geodesic radial motion. The newly found radially accelerated trajectories are generalizations of the known uniformly accelerated stationary observers in the anti-de Sitter universe.

In the last part of our work we have focused on the return journey of the rocketeer. It had been suggested that having uniform deceleration would be enough in order to have an actual return journey \cite{rindler1960hyperbolic}. Here we have proved that even if this condition is necessary, it is not sufficient for all spacetimes. In particular, among the cosmological models analyzed here, in the de Sitter case Eq.~\eqref{bound} must be satisfied for a return journey to be possible. 

\begin{acknowledgements}
I am grateful to Georgios Loukes-Gerakopoulos, Giovanni Acquaviva and Ji\v{r}\'{i} Bi{\v{c}}{\'a}k for valuable discussions on this work and useful comments on the manuscript. 
\end{acknowledgements}

\bibliographystyle{unsrt}
\bibliography{ref}

\end{document}